\documentclass{mem}

\usepackage{natbib}
\usepackage{txfonts}
\usepackage{balance}
\usepackage{graphicx}
\usepackage[a4paper]{hyperref}
\idline{00}{29}

\newcommand\omp{\Omega_{\rm p}}
\newcommand\omtw{\Omega}
\newcommand\omint{\Omega_{\rm 0}}
\newcommand\kin{{\cal V}}
\newcommand\pin{{\cal X}}

\newcommand\DLwidth{{$w_{\rm dl}$}}
\newcommand\DLslope{{$s_{\rm dl}$}}
\newcommand\DLxmin{{$x_{\rm min}$}}

\begin{document}

\title{The effect of dust on the Tremaine-Weinberg method}

\subtitle{}

\author{J. Gerssen\inst{1} \and V. P. Debattista\inst{2}}

\offprints{Joris Gerssen; \email{jgerssen@aip.de}}
 
\institute{Astrophysikalischer Institut Potsdam, Potsdam, Germany
\and
  Centre for Astrophysics, University of Central Lancashire, Preston,
  UK}

\authorrunning{Gerssen and Debattista}

\titlerunning{Dust effect on Tremaine-Weinberg measurements}

\abstract{We investigate the effect of dust on the observed rotation rate of a stellar bar as measured by application of the Tremaine \& Weinberg (TW) method.  As the method requires that the tracer satisfies the continuity equation, it has been applied largely to early-type barred galaxies.  We show using numerical simulations of barred galaxies that dust attenuation factors typically found in these systems change the observed bar pattern speed by 20--40\%. We also address the effect of star formation on the TW method and find that it does not change the results significantly.  This suggests that applications of the TW method can be extended to include barred galaxies covering the full range of Hubble type.
}

\maketitle{}

\section{Introduction}

The rate at which a bar rotates, its pattern speed, $\omp$, is the
principle parameter controlling a barred (SB) galaxy's dynamics and
morphology.  Most determinations of this parameter are indirect. An
often used method is to match hydrodynamical simulations of SB
galaxies to observed velocity fields.  The only direct and
model-independent technique to measure $\omp$ is the Tremaine \&
Weinberg (1984, hereafter TW) method.  They show that for any tracer
that satisfies the continuity equation, $\kin = \pin\,\omp\,\sin{i}$.
Here, $\pin \equiv \int h(Y)\,X\,\Sigma\,dX\,dY$ and $\kin \equiv \int
h(Y)\,V_{\rm los}\,\Sigma\,dX\,dY$ are the luminosity-weighted mean
positions and velocities respectively.

The number of successful applications of the TW method is rather
limited and confined mostly to early type SB galaxies (e.g.,
\citealt{agu02}; \citealt{ger03}). Late-type barred galaxies often
display prominent dust lanes along the leading edges of the bar which
can invalidate the assumption of tracer-continuity. Our motivation for
this study is to explore how important dust is for TW measurements of
both early and especially late-type galaxies.  In the latter, star
formation can be a potential problem as it violates the basic
assumption of the TW method.  We therefore also explore the ability to
measure $\omp$ in a hydrodynamical simulation of a barred galaxy that
includes star formation \citep{deb06}.

\begin{figure*}[t!]
\resizebox{\hsize}{!}{\includegraphics[clip=true]{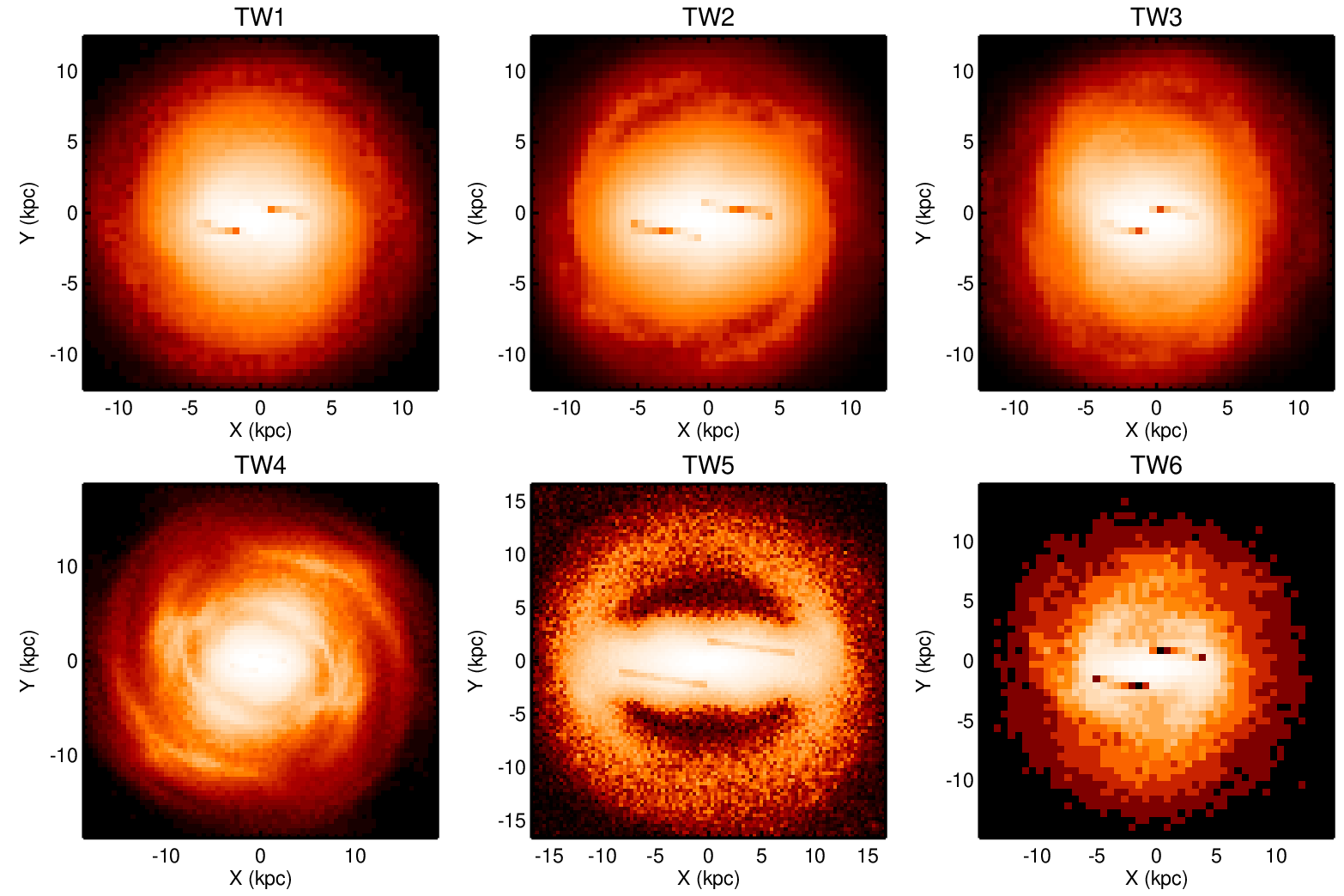}}
\caption{\footnotesize 
  Examples of the projected distribution of particle weights in our
  models. The models are published in \citet[][TW1-TW4]{deb03},
  \citet[][TW5]{deb00}, and \citet[][TW6]{deb06}.  They are shown here
  face-on to highlight their morphological features. Dust lanes have
  been included to illustrate how dust attenuation is implemented in
  our simulations.  For clarity the dust lanes have an exaggerated
  extinction ($A_V = 15$) and width.}
\label{f:maps}
\end{figure*}

\section{Method}

In a particle implementation of the TW method, the integrals over
surface brightness and velocities are replaced by sums over the
particles.  Five out of our six N-body models all have particles that
are coeval and have equal mass. We therefore start by assigning the
same intrinsic weights to all particles in these models.  Dust lanes
on the leading edges of the bars are implemented by lowering the
weights of particles that reside in these areas.  We assume that the
dust distribution within the dust lanes is described by a double
exponential model.  That is, $D_{0,{\rm lane}} = e^{-R/h_{R,{\rm
lane}}}\,e^{-|z|/h_{z,{\rm lane}}}$ when a particle resides in the dust
lane and is $0$ otherwise.  Particles that are shadowed by dust lanes
contribute to the TW integrals with lower weights.  Unlike a
foreground screen model, the dust extinction in our models varies with
position within the disc.  For each particle we need to calculate the
amount of intervening dust by integrating the projected dust
distribution along the line-of-sight.  The weight for each particle is
then given by $w_i = e^{-\tau_i}$, where the optical depth $\tau_i$ is
obtained from $\tau_i = \int^{s_i}_{-\infty} D(s')\,ds'$.  Examples
of the weight distribution in each of our six models are shown in
Fig.~\ref{f:maps}. They are shown here face-on to better illustrate
their morphological features.

\begin{figure*}[]
\resizebox{\hsize}{!}{\includegraphics[clip=true]{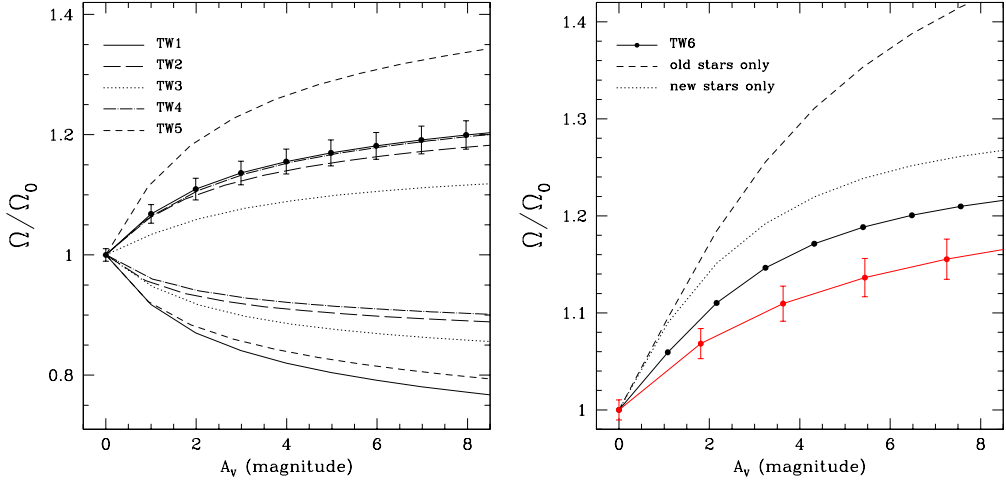}}
\caption{ \footnotesize 
  Left panel: The ratio $\omtw / \omint$ of the bar pattern speed
  observed with the TW method to the intrinsic bar pattern speed as a
  function of dust lane extinction $A_V$.  The different curves show
  the behaviour of this ratio for the different N-body models.  Two
  curves are shown for each model.  Curves with a ratio $> 1$ are at
  $\rm PA_{\rm bar} = +45^\circ$, while curves with a ratio $< 1$ are
  at $\rm PA_{\rm bar} = -45^\circ$. The dust lane geometry and dust
  distribution are the same for every model: \DLwidth/$a_{\rm B} =
  0.1$, \DLslope\ = $-0.3$, \DLxmin/$a_{\rm B} = -0.1$, $h_{R,{\rm
  lane}} = 10$ and $h_{z,{\rm lane}} = 0.1$.  For clarity, the
  $1\sigma$ errors are shown for model TW1 only.  Right panel: The bar
  pattern speed ratio as a function of dust extinction for the model
  that includes star formation (TW6).  The result is not very
  different from the models shown in the left panel. For comparison
  model TW1 is overplotted as the solid line with error bars.}
\label{f:omega}
\end{figure*}

\section{Results}

\subsection{Dust extinction}

In these proceedings we explore the behaviour of the measured pattern
speed as a function of dust extinction, expressed in $A_V$ magnitude
($A_V = 1.086\,\tau_0$).  The effects of adding a global dust disc and
varying the dust lane geometry are small and are detailed in
\citep{ger07}. Fig.~\ref{f:omega} (left panel) shows the ratio
$\Omega_{\rm p}/\Omega_0$ of the bar pattern speed observed with the
TW method to the intrinsic bar pattern speed as a function of dust
lane extinction $A_V$ (defined face-on). The different curves show the
behaviour of this ratio for the different N-body models (TW1 to TW5).
Two curves are shown for each model.  Curves with a ratio $>1$ are at
$\rm PA_{\rm bar} = +45^{\circ}$, while curves with a ratio $<1$ are
at $\rm PA_{\rm bar} = -45^{\circ}$.  The dust lane geometry and dust
distribution are the same for every model.

\subsection{Star formation}

In late type systems, vigorous star formation may invalidate the TW
assumption that the tracer population satisfies the continuum
condition.  We use an N-body + SPH model (TW6) to explore these
effects. The results, shown on the left, are qualitatively similar to
the results derived from the N-body models TW1 to TW5. For a
quantitatively comparison model TW1 is overplotted as the solid line
with error bars in Fig.~\ref{f:omega} (right panel).

\section{Conclusions}

With realistic amounts of absorption in the dust lanes ($A_V \sim 3$)
the models predict observationally insignificant errors (20--40\%) in
the observed pattern speed.  Our results suggest that the application
of the TW method can be extended to later-type barred galaxies and
facilitate direct comparisons with pattern speeds measured by
hydrodynamical modeling.


\bibliographystyle{aa}

\begin{thebibliography}{}

\bibitem[{Aguerri et al.(2002)}]{agu02} 
  Aguerri, J. A. L., Debattista, V. P., \& Corsini, E. M. 2003, MNRAS,
  338, 465

\bibitem[{Debattista(2003)}]{deb03} 
  Debattista, V. P. 2003, MNRAS, 342, 1194 

\bibitem[{Debattista \& Sellwood(2000)}]{deb00} 
  Debattista, V. P., \& Sellwood, J. A. 2000, ApJ, 543, 704 

\bibitem[{Debattista et al.(2006)}]{deb06} 
  Debattista, V. P., Mayer, L., Carollo, C. M., Moore, B., Wadsley,
  J., \& Quinn, T. 2006 ApJ, 645, 209

\bibitem[{Gerssen \& Debattista(2007)}]{ger07} 
  Gerssen, J., \& Debattista, V. P. 2007, MNRAS, 378, 189

\bibitem[{Gerssen et al.(2003)}]{ger03} 
  Gerssen, J., Kuijken, K., \& Merrifield, M. R. 2003, MNRAS, 345, 261

\bibitem[{Tremaine \& Weinberg(1984)}]{tw84} 
  Tremaine, S., \& Weinberg, M. D. 1984, ApJ, 282, L5 (TW)

\end{thebibliography}

\end{document}